\documentclass[12pt]{article}
\usepackage{graphicx} % Required for inserting images
\usepackage{setspace}
\usepackage{subcaption}
\usepackage{tikz-cd}
\usepackage{float}
\usepackage{amsmath}
\usepackage{tocloft}
\usepackage{hyperref}
\newcommand{\al}{\alpha}
\newcommand{\be}{\beta}
\newcommand{\ga}{\gamma}

\renewenvironment{abstract}
 {\large
  \begin{center}
  \bfseries Abstract
  \end{center}
  \quotation}
 {\endquotation}

\title{Fuzzy-Space Engineering}
\author{Paul Schreivogl\textsuperscript{1}, Richard Schweiger\textsuperscript{2}}

\date{
    \small
    FH Technikum Wien Höchstädtplatz 6 1200 Vienna, Austria\\
    NM-8, Dorotheergasse 10, 1010 Vienna, Austria\\
    March 2024
}
\begin{document}
%\begin{spacing}{1.2}
\maketitle

\begin{abstract}
    \small
\begin{spacing}{1.2}
    The techniques developed for matrix models and fuzzy geometry are powerful tools for representing strings and membranes in quantum physics. We study the representation of fuzzy surfaces using these techniques. This involves constructing graphs and writing their coordinates and connectivity into matrices. To construct arbitrary graphs and quickly change them, we use 3D software. A script generates the three matrices from the graphs. These matrices are then processed in Wolfram Mathematica to calculate the zero modes of the Dirac operator. Our first result shows the quantization of a two-dimensional Trefoil knot. Additional examples illustrate various properties and behaviors of this process. This helps us to gain a deeper understanding of fuzzy spaces and zero-mode surfaces. This work contributes to advancing the understanding of visualization aspects in fuzzy geometry.
\end{spacing}    
\end{abstract}

\noindent \textbf{Keywords}: Fuzzy geometry, zero mode surface, fuzzy Trefoil knot, Dirac operator, non-commutative geometry, matrix theory

\vfill
\textsuperscript{1} schreivo@technikum-wien.at,
\textsuperscript{2} office@nm-8.com
\clearpage
 \tableofcontents{\large}
\clearpage
\section{Introduction}
Matrix models, noncommutative geometry, and fuzzy geometry provide exciting mathematical tools for representing strings, membranes, and quantum physics \cite{Ishibashi:1996xs},\cite{Banks:1996vh},\cite{Connes:1998}. For reviews, see \cite{Taylor:2001vb} and \cite{Ydri:2017ncg}. A very promising tool for the investigation of matrix models \cite{Berenstein:2012ts}, \cite{Ishiki:2015saa} are coherent states \cite{Perelomov:1972}. These states have led to zero-mode surfaces of the Dirac operator, with which one can represent fuzzy or quantized spaces \cite{deBadyn:2015sca},\cite{Sykora:2016ldn}, \cite{Schneiderbauer:2016wub}.
 At the beginning of the development of fuzzy geometries, very symmetric spaces were studied. For example, the fuzzy torus \cite{hoppe1982} or the fuzzy sphere \cite{madore1992}.Embedding graphs into matrices allows for the description and plotting of more complex fuzzy surfaces. This method also enables the treatment of compact Riemann surfaces with a higher genus.\\
 In this work, we focus on geometrical properties and the behavior of this procedure, rather than on physical sensible solutions for matrix models. We will use the above-mentioned methods and further investigate their behavior, enabling a more adept construction of sophisticated graphs to plot zero mode surfaces. \\We manually created custom graphs using the 3D application Blender $\footnote{Blender Foundation. (2023). Blender, Version 3.0. Amsterdam, Netherlands}$. A Python script then extracted these models and encoded the graph data into matrices. These matrices describe the coordinates of the graph's nodes and edges. Our Python script writes all necessary commands for plotting and copies them into the clipboard, to directly paste them in a Wolfram Mathematica notebook
$\footnote{Wolfram Research, Inc. (2022). Mathematica, version 13.0. Champaign, IL.}$. \\
This script creates a swift bridge between intuitive modeling of graphs in Blender and fast plotting of their zero-mode surfaces in Wolfram Mathematica. In this way, we can produce and examine plots of complex graphs. Through this workflow, we successfully quantized a two-dimensional Trefoil knot. \\
%We studied topological properties like the separation of one zero mode surface into two.
We provide a dynamic example where one zero-mode surface separates into two zero mode surfaces.
We have studied a simple example of deformations where we examine three aspects: the deformation of coordinates of a fuzzy space, where the deformations are  associated with a non-commutative gauge field, and how the deformation of coordinates acts on the zero mode surface, as well as how the non-commutative gauge field affects the graphs.
\clearpage
\section{Matrices, zero-surfaces and graphs}
In this section we provide a concise summary of the mathematical approach used in the paper, as will be referenced in subsequent sections.
For a precise and detailed description of the correlation between coherent states ( and quasicoherent states), zero-mode surfaces, and strings, refer to \cite{Sykora:2016ldn} and \cite{Schneiderbauer:2016wub}. Subsequently, we will further explore the details outlined in \cite{Sykora:2016ldn}, as it specifically addresses the assignment of graphs to two-dimensional compact surfaces.
\\
We define the extended coordinate function on a background space 
\begin{eqnarray}
Y^{a} = \left( \begin{array}{cc}
   X^{a} &  \\
     &  x^{a}\\
       \end{array} \right),
\end{eqnarray}
where the hermitian $N\times N$-matrices $X^{a}$ with $a=1,...,d$, fulfill the commutation relations 
 \begin{eqnarray}
[X^a,X^b]=\theta(X)^{ab}.
\end{eqnarray}
The coordinate functions $Y^{a}$ represent a fuzzy space together with a point probe at $x^{a} \in \mathbf {R}^{d}$.
We consider a spinor valued matrix  $\Psi$ for the Dirac operator on the background $\mid\psi\rangle$ which is  a spinor valued vector
\begin{eqnarray}
\Psi = \left( \begin{array}{cc}
    &  \mid\psi\rangle \\
      \langle \psi\mid   &  \\
       \end{array} \right).
\end{eqnarray}
The localized Dirac operator reads then
\begin{eqnarray}
D_{Y}=\sum_a\gamma_{a}[Y^{a},\Psi] = \left( \begin{array}{cc}
    &  D_{x}\mid\psi\rangle \\
     \langle \psi\mid D_{x}  &  \\
       \end{array} \right),
\end{eqnarray}
with $\gamma_{a}$ the generators of the Clifford algebra. The operator $D_{x}$ is written as
\begin{eqnarray}
D_{x}=\sum_{a}\gamma_{a}(X^{a}-x^{a}\mathbf{1}).
\end{eqnarray}
In order to find zero modes of the Dirac operator we have to analyze this equation
\begin{eqnarray}
D_{x} \mid\psi\rangle =0  
\end{eqnarray}
or equivalently compute the  determinant
\begin{eqnarray}\label{zeromode}
 det(\sum_{a}\gamma_{a}(X^{a}-x^{a}\mathbf{1}) ) = P(x^{a}) = 0.
\end{eqnarray}
This equation defines a multivariate polynomial $P(x^{a})$ in $d$ variables. The polynomial defines a so-called "zero-mode manifold".\\
\\
We list some properties of the determinant (\ref{zeromode})
     \begin{itemize}
         \item Invariance under unitary transformation
         \begin{eqnarray}\label{trans}
          X^{a}\rightarrow U X^{a} U^{\dagger},
         \end{eqnarray}
         which can be interpreted as a non-commutative symplectic coordinate transformation.
         \item Invariance under translations, rotations, and scaling \begin{eqnarray}
          X^{a}\rightarrow \alpha R_{b}^{a}X^{b} +c^{a}\quad \text{and} \quad x^{a}\rightarrow \alpha R_{b}^{a}x^{b} +c^{a} 
         \end{eqnarray}
         \item In the case of block diagonal matrices, 
         there is a splitting of the fuzzy space into a direct sum of smaller fuzzy spaces
         \begin{eqnarray}
   X^{a} = \left( \begin{array}{cc}
   X^{a}_{1} &  \\
     &  X^{a}_{2}\\
       \end{array} \right)
\end{eqnarray}
       and the determinant factors in a product
      \begin{eqnarray}
     det D_{x}=  det(\gamma_{a}(X^{a}_{1}-x^{a}\mathbf{1}) ) det(\gamma_{a}(X^{a}_{2}-x^{a}\mathbf{1}) )  =0 \nonumber
\end{eqnarray} 
so 
\begin{eqnarray}
 det(\gamma_{a}(X^{a}_{1}-x^{a}\mathbf{1}) ) =0 \quad \text{or} \quad det(\gamma_{a}(X^{a}_{2}-x^{a}\mathbf{1}) )  =0.\nonumber
\end{eqnarray} 
\end{itemize}
We can rewrite equation ($\ref{zeromode}$) in three dimensions in terms of the matrices $X$,$Y$ and $Z$. The Gamma matrices $\gamma_a$ are the three Pauli matrices and the Dirac operator reads
\begin{eqnarray}\label{VZ}
&&det \left( \begin{array}{cc}
   Z-z&  (X-x)-i(Y-y)\\
    (X-x)+i(Y-y) & -(Z-z)\\
       \end{array} \right)\\
       &=&
det \left( \begin{array}{cc}
     Z-z& V^{\dagger} -\bar{v}\\
     V-v &  -(Z-z)\\
       \end{array} \right)=0,
\end{eqnarray}

where $V=X+iY$ and $v=x+iy$.
\subsection{Mapping graphs to matrices}
%\subsection{Basic building block and general mapping}
We repeat the procedure from the literature \cite{Sykora:2016ldn} on how to encode a graph into matrices $X$, $Y$, $Z$. Consider the simple case of a graph with two vertices located at $(x_{1},y_{1},z_{1})$ and $(x_{2},y_{2},z_{2})$ in  $\mathbf{R}^{3}$, which are connected by an edge. Note that the resulting graph is a directed graph. The edge is labeled with two numbers $s_x$ and $s_y$, with $s_x$ being purely real and $s_y$ purely imaginary. The graph is mapped to matrices as follows...
\begin{eqnarray}
X = \left( \begin{array}{cc}
   x_{1} & s_{x} \\
     s_{x}&  x_{2}\\
       \end{array} \right),
       Y = \left( \begin{array}{cc}
   y_{_1} & s_{y} \\
     \bar{s}_y&  y_{2}\\
       \end{array} \right),
       Z = \left( \begin{array}{cc}
   z_{1} & 0 \\
     0&  z_{2}\\
       \end{array} \right).
\end{eqnarray}
Next, the hermitian matrices $X$ and $Y$ are combined to one complex matrix $V=X+iY$ and  two complex numbers $v_{1}=x_{1}+iy_{1}$ and $v_{2}=x_{2}+iy_{2}$ and $s_{12}=s_{x}+s_{y}$ and $s_{21}=s_{x}+\bar{s_{y}}$. Then the matrices in
($\ref{VZ}$) become
\begin{eqnarray}
V = \left( \begin{array}{cc}
   v_{1} & s_{12} \\
     s_{21}&  v_{2}\\
       \end{array} \right),\quad
       Z = \left( \begin{array}{cc}
   z_{1} & 0 \\
     0&  z_{2}\\
       \end{array} \right).
\end{eqnarray}
\\
 We fix the positions of the points $\vec{x_{1}}=(0,0,-1/2)$, $\vec{x_{2}}=(0,0,1/2)$ and the radii to $s_{x}=1/2$ and $s_{y}=-i/2$,
then the matrices $X$, $Y$ and $Z$ become proportional to the Pauli matrices $\sigma_{1}$, $\sigma_{2}$ and $\sigma_{3}$ and define the smallest fuzzy sphere
\begin{eqnarray}
X = \left( \begin{array}{cc}
   0 & \frac{1}{2} \\
     \frac{1}{2}&  0\\
       \end{array} \right)=\frac{\sigma_{1}}{2},
       Y = \left( \begin{array}{cc}
   0& -\frac{i}{2} \\
    \frac{i}{2}&  0\\
       \end{array} \right)=\frac{\sigma_{2}}{2},
       Z = \left( \begin{array}{cc}
  \frac{1}{2} & 0 \\
     0& -\frac{1}{2}\\
       \end{array} \right)=\frac{\sigma_{3}}{2},
\end{eqnarray}
with the commutation relation $[X,Y]=iZ$ and with the Casimir operator $X^2+Y^2+Z^2=3/4$.
In this case the $V$ matrix becomes
\begin{eqnarray}
V= X+iY= \left( \begin{array}{cc}
   0 & 1 \\
     0&  0\\
       \end{array} \right),\quad
       Z = \frac{1}{2}\left( \begin{array}{cc}
  1& 0 \\
     0&  -1\\
       \end{array} \right).
\end{eqnarray}
Directly evaluating the determinant in equation ($\ref{zeromode}$) using the matrices $X$, $Y$ and $Z$ yields the following result for the zero mode surface.
\begin{eqnarray}
det(\gamma_{a}(X^{a}-x^{a}\mathbf{1}) )=(x^2+y^2+z^2-1/4)(x^2+y^2+z^2+3/4)=0,\nonumber\\
\end{eqnarray}
which defines the sphere $x^2+y^2+z^2-1/4=0$.\\

A physical interpretation of this construction, is that a closed string is created at the point $\vec{x_{1}}=(0,0,-1/2)$ and disappears at the point $\vec{x_{2}}=(0,0,1/2)$. The two real numbers $s_x$ and $s_y$ are then interpreted as the maxima of the string radii.\\

We summarize the procedure to produce the zero mode surfaces by the sequence
\begin{equation}
\text{Graph} \longrightarrow X^a \longrightarrow \text{det}(\sum_{a}\gamma_{a}(X^{a}-x^{a}\mathbf{1})) = P(x^a) = 0.
\end{equation}

\paragraph{General graphs}
Consider a simple graph that has no edges and consists of $N$ nodes $(x_{i},y_{i},z_{i})$. The determinant $detD_{x}$ is invariant under the permutation of node coordinates. These permutations form a subgroup of transformations that leave the determinant invariant.
\begin{figure}[H]
  \centering
  \begin{subfigure}{0.48\textwidth}
    \centering
    \includegraphics[width=0.8\linewidth]{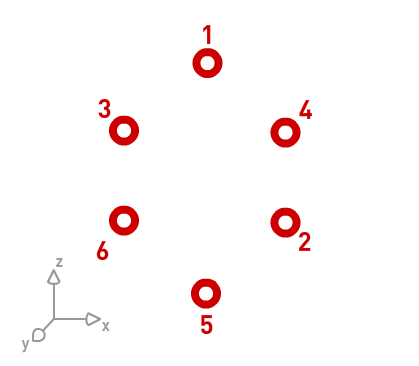}
    \caption{Nodes without connecting edges.}
    \label{fig:subfig1}
  \end{subfigure}
  \hfill
  \begin{subfigure}{0.48\textwidth}
    \centering
    \includegraphics[width=0.8\linewidth]{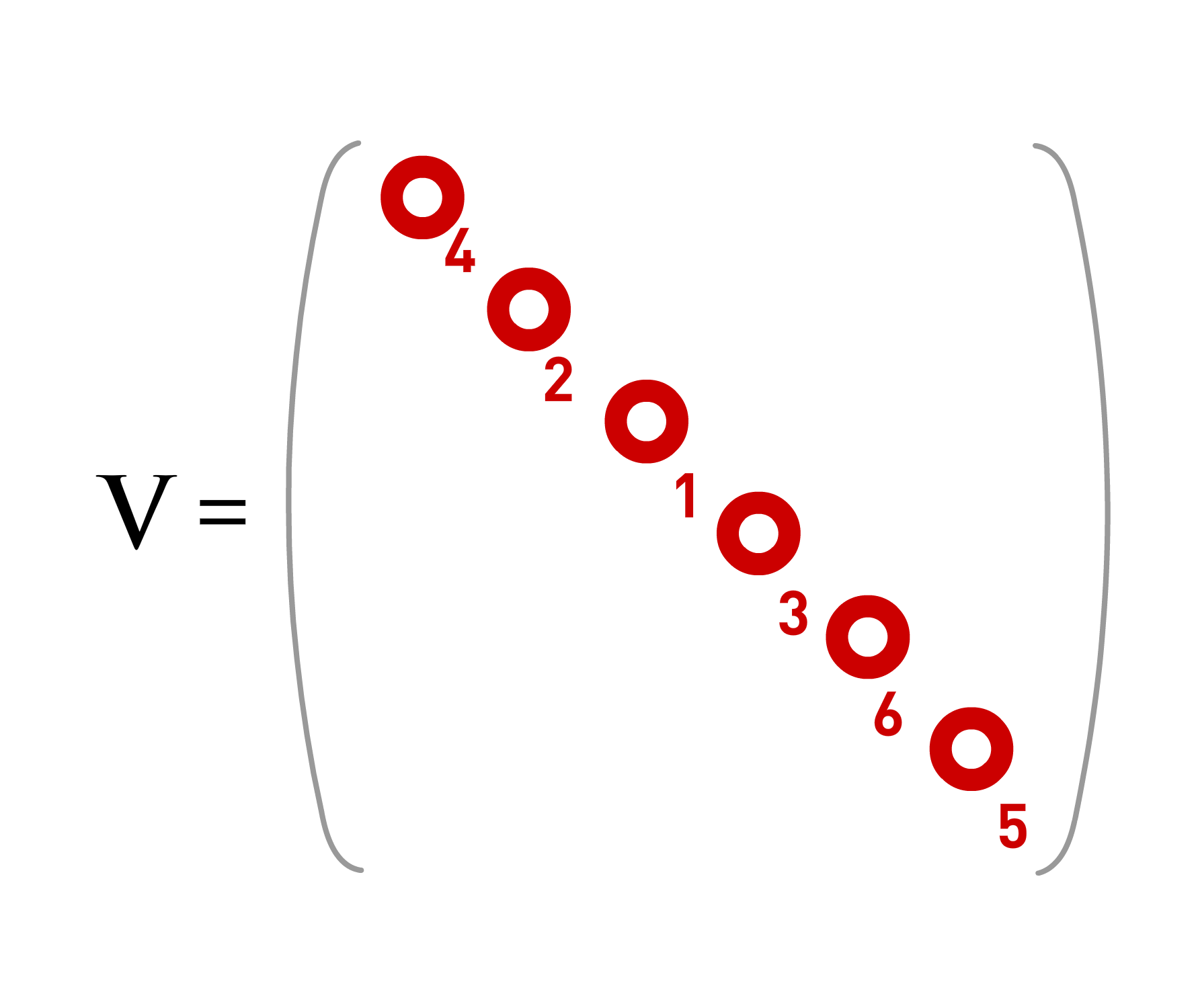}
    \caption{No particular order of nodes necessary.}
    \label{fig:subfig2}
  \end{subfigure}
  \caption{Set of unconnected nodes.}
  \label{fig:both_images}
\end{figure}

\begin{figure}[H]
  \centering
  \begin{subfigure}{0.48\textwidth}
    \centering
    \includegraphics[width=0.8\linewidth]{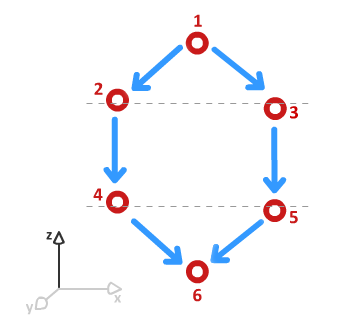}
    \caption{Arrangement of nodes and edges.}
    \label{fig:subfig1}
  \end{subfigure}
  \hfill
  \begin{subfigure}{0.48\textwidth}
    \centering
    \includegraphics[width=0.8\linewidth]{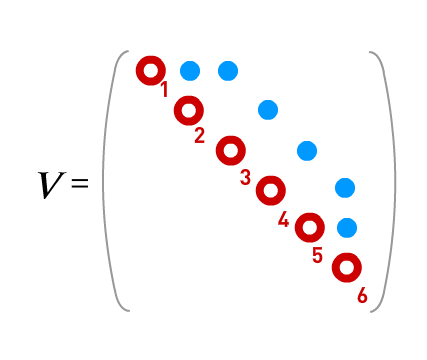}
    \caption{Increasing z coordinate defines the order in $V$.}
    \label{fig:subfig2}
  \end{subfigure}
  \caption{Connected nodes.}
  \label{fig:both_images2}
\end{figure}

In order to include edges between points, we introduce off-diagonal values in the matrices, as seen in the example of the Fuzzy sphere's $X$ and $Y$ matrices above. \\ If there is an edge between two points, we label that edge with two real values $(s_{xij} , s_{yij})$, where $i$ and $j$ are the indices of the nodes being connected. This describes a directed edge, and encode it in the three $N\times N$ matrices $X$, $Y$.\\
graphs.
\clearpage
To summarize the whole process of reading geometric data of a graph and constructing matrices from it, the following list of properties are obligatory:
\\
\begin{itemize}
    \item We identify the diagonals of the matrices with the node coordinates
    \begin{eqnarray}
        X_{ii}&=&x_{i}\nonumber\\
        Y_{ii}&=&y_{i}\\
        Z_{ii}&=&z_{i}\nonumber
    \end{eqnarray}
    \item If there is an edge between vertices i and j, then there are entries in the off-diagonal elements
    \begin{eqnarray}
        X_{ij}&=&s_{x_{ij}}\nonumber\\
        X_{ji}&=&s_{x_{ij}}\nonumber\\
        Y_{ij}&=&s_{y_{ij}}\\
        Y_{ji}&=&-s_{y_{ij}}\nonumber
    \end{eqnarray}        
     \item  All other entries in the matrices are zero.
\end{itemize}
\vspace{10mm}
 By using the complexification of the the matrix $X$ and $Y$, the $V$ matrix can be written as
 \begin{eqnarray}
        Z_{ii}&=&z_{i} \nonumber\\
        V_{ii}&=& x_{i}+iy_{i}=v_{i}\nonumber\\
        V_{ij}&=& s_{ij}=s_{x_{ij}}- i s_{y_{ij}}\\
        V_{ji}&=&s_{ji}=s_{x_{ij}}+i s_{y_{ij}}\nonumber
  \end{eqnarray}
  \\
 Changing the direction of a single edge from $i$ to $j$ corresponds to exchanging $s_{ij}$ with $s_{ji}$ and does not give the same zero mode surface. The next paragraph focuses on the behaviour of zero mode surfaces, when the edge directions are changed in this way.
 \clearpage
\paragraph{Spatial properties} Nodes that are connected with edges, as shown in Figure $\ref{fig:both_images2}$, can be arranged in $V$ with descending $z$ component to get a tidy cells distribution. Especially if the $s_y = -i s_x$. Then below the diagonal they add up, and above they subtract to 0.
The nodes are sorted with descending z coordinate, then the edges are automatically directed along the negative direction of the z axis. 
\paragraph{Edge direction and radii}
The direction of an edge means, it points from one node $i$ to another node $j$, with $i \neq j$. 
\begin{figure}[H]
  \centering
  \begin{subfigure}{0.48\textwidth}
    \centering
    \includegraphics[width=0.8\linewidth]{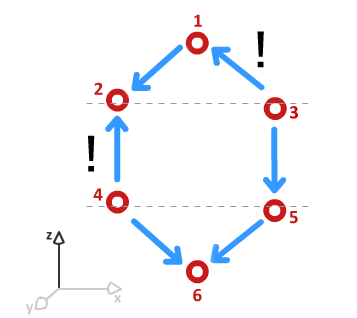}
    \caption{Altered direction of edges.}
    \label{fig:subfig1}
  \end{subfigure}
  \hfill
  \begin{subfigure}{0.48\textwidth}
    \centering
    \includegraphics[width=0.8\linewidth]{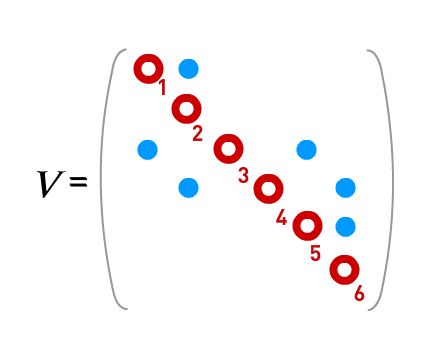}
    \caption{reordered radii.}
    \label{fig:subfig2}
  \end{subfigure}
  \caption{Connected nodes with altered directions}
  \label{fig:both_alteredimages3}
\end{figure}

\begin{figure}[H]
    \centering
    \includegraphics[width=0.5\linewidth]{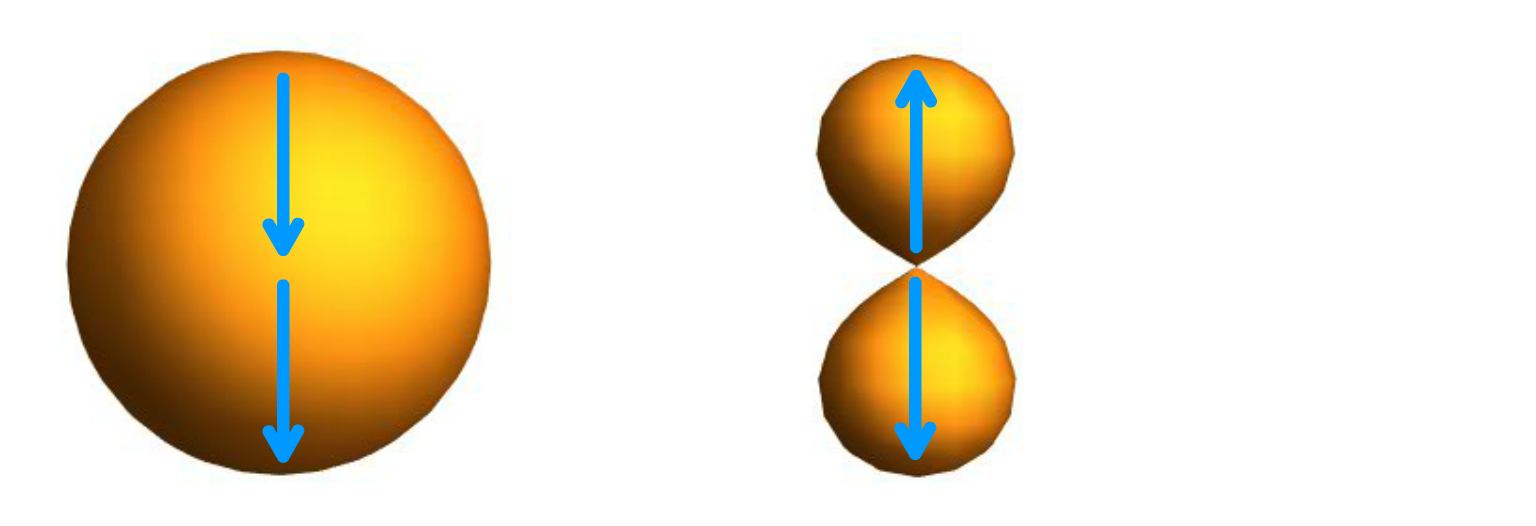}
    \caption{The zero mode surface changes, if an edges direction is reversed.}
    \label{fig:edgedirection}
\end{figure}

The edges shown in Figure $\ref{fig:both_images2}$ are the blue dots in cells, where the $s_{ij}$ are non zero. The change in direction, results in the transposed cells $s_{ji}$ to turn out nonzero, as shown in Figure $\ref{fig:both_alteredimages3}$.
\\
The direction of an edge is crucial for the zero mode surface. If we reverse one edge in a two edge fuzzy sphere the opposing edges result in a zero mode surface thats not a sphere, but rather resembles an hourglass. See Figure $\ref{fig:edgedirection}$.

If an edge's z-component gets too small, compared to its fuzzy radius $s_{ij}$, the zero mode surface results in a gap between the nodes.
\begin{figure}[H]
    \centering
    \includegraphics[width=1\linewidth]{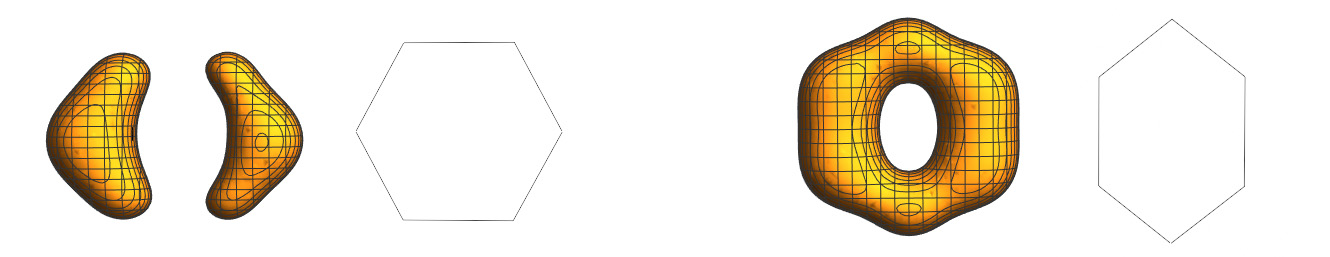}
    \caption{Torus with and without horizontal edges.}
    \label{fig:tori_compare}
\end{figure}
As shown in Figure $\ref{fig:tori_compare}$, such horizontal edges won't create a corresponding zero mode surface. The following graphs will avoid these horizontal edges, for a consistent correspondence between the zero mode surfaces and the graphs. This behaviour is examined further in section 3.2.

\section{Modeled graphs}
We use the open-source 3D software Blender, in order to quickly model arbitrary 3D-graphs. In Blender objects have a suitable anatomy, as they consist of vertices, edges and faces. We can use vertices and edges, to represent the nodes and directed edges of graphs. To plot a zero mode surface in Mathematica, we wrote a Python script that reads a 3D-graph in Blender, and writes the determinant in Mathematica syntax to the clipboard.\\ 
The script rearranges the  vertices and edges of the graph mesh, to ensure the demanded structure shown in Figure $\ref{fig:both_images2}$. This resulted in a swift and convenient workflow from Blender to Mathematica.\\
The flowchart of the Python script to generate Mathematica code is shown in section 6 together with the link to download it.
\subsection{Fuzzy Trefoil knot}
\begin{figure}[H]
    \centering
    \includegraphics[width=0.6\linewidth]{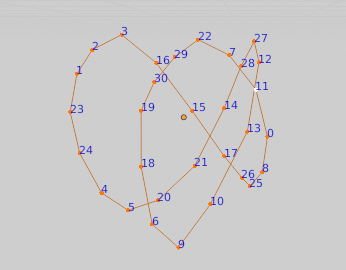}
    \caption{Trefoil knot graph.}
    \label{fig:trefoil_mesh}
\end{figure}

\begin{figure}[H]
  \centering
  \begin{subfigure}{0.48\textwidth}
    \centering
    \includegraphics[width=0.8\linewidth]{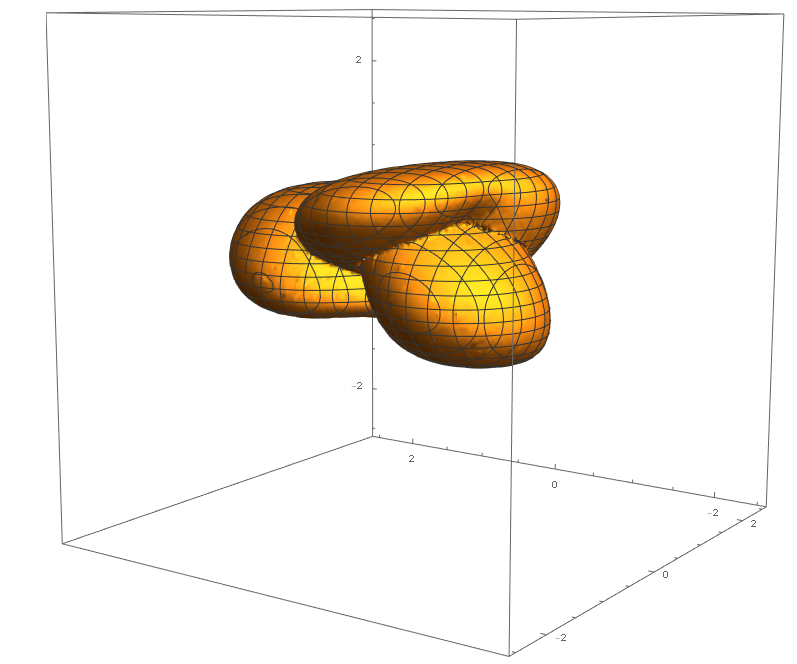}
    \caption{Thicker volume → intersections}
    \label{fig:subfig1}
  \end{subfigure}
  \hfill
  \begin{subfigure}{0.48\textwidth}
    \centering
    \includegraphics[width=0.8\linewidth]{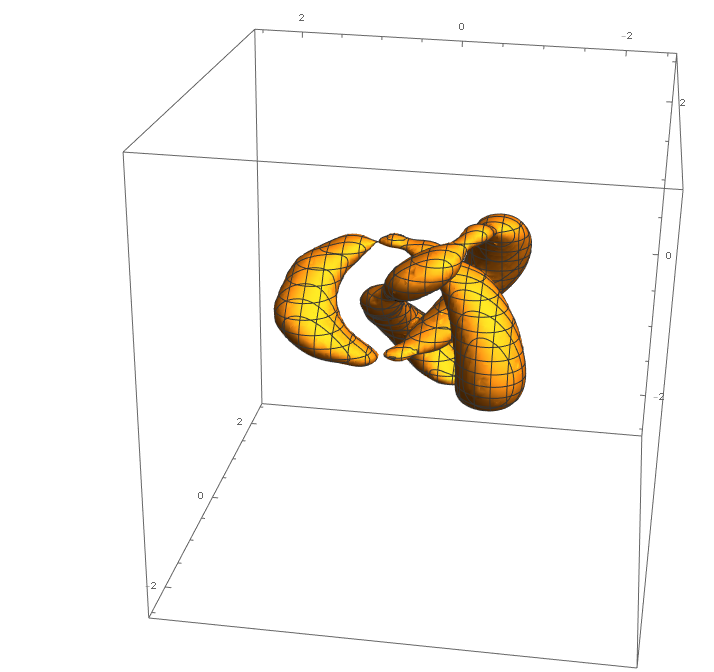}
    \caption{Thinner volume → separations}
    \label{fig:subfig2}
  \end{subfigure}
  \caption{Different radii $s_{xy}$ and vertices counts shaping the fuzzy surface.}
  \label{fig:intersect-seperate}
\end{figure}

\begin{figure}[H]
    \centering
    \includegraphics[width=0.7\linewidth]{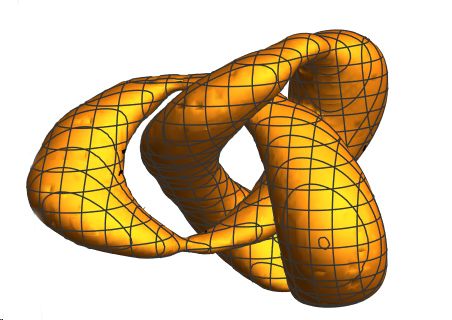}
    \caption{Fuzzy Trefoil knot.}
    \label{fig:TREFOIL}
\end{figure}

With the above presented method, we construct a  Trefoil knot as 3D-graph and plot a zero mode surface from it. The plot represents a two dimensional Trefoil knot embedded in $\mathbf{R}^{3}$. We needed to balance the number of vertices in our 3D-graph. Too few vertices would not capture the geometry accurately. Too many could cause the plot to fail, as the degree of the determinant’s polynomial grows with the matrix size.\\
We found the Trefoil knot with least vertices, to be a graph with 30 nodes and edges as shown in Figure \ref{fig:trefoil_mesh}.
\\
The edges had to be kept shorter than the distance between the separate parts of the mesh. Otherwise, the parts would intersect. If fewer vertices are used to form the knot, the edge lengths increase until the fuzzy surface separates into distinct parts. Figure \ref{fig:intersect-seperate} shows two examples for such intersection and separation.
Sophisticated fuzzy surfaces have a lower limit of the least nodes in their graph. This lower limit opposes an upper limit of nodes to solve the determinant $det(D_x)=0$ in Mathematica. \\ The resulting 30-node Trefoil knot is shown in Figure \ref{fig:TREFOIL}. Download the corresponding notebook here \cite{SchSchM1:2024}.

\subsection{Volumetric representation}
\begin{figure}[H]
    \centering
    \includegraphics[width=1.0\linewidth]{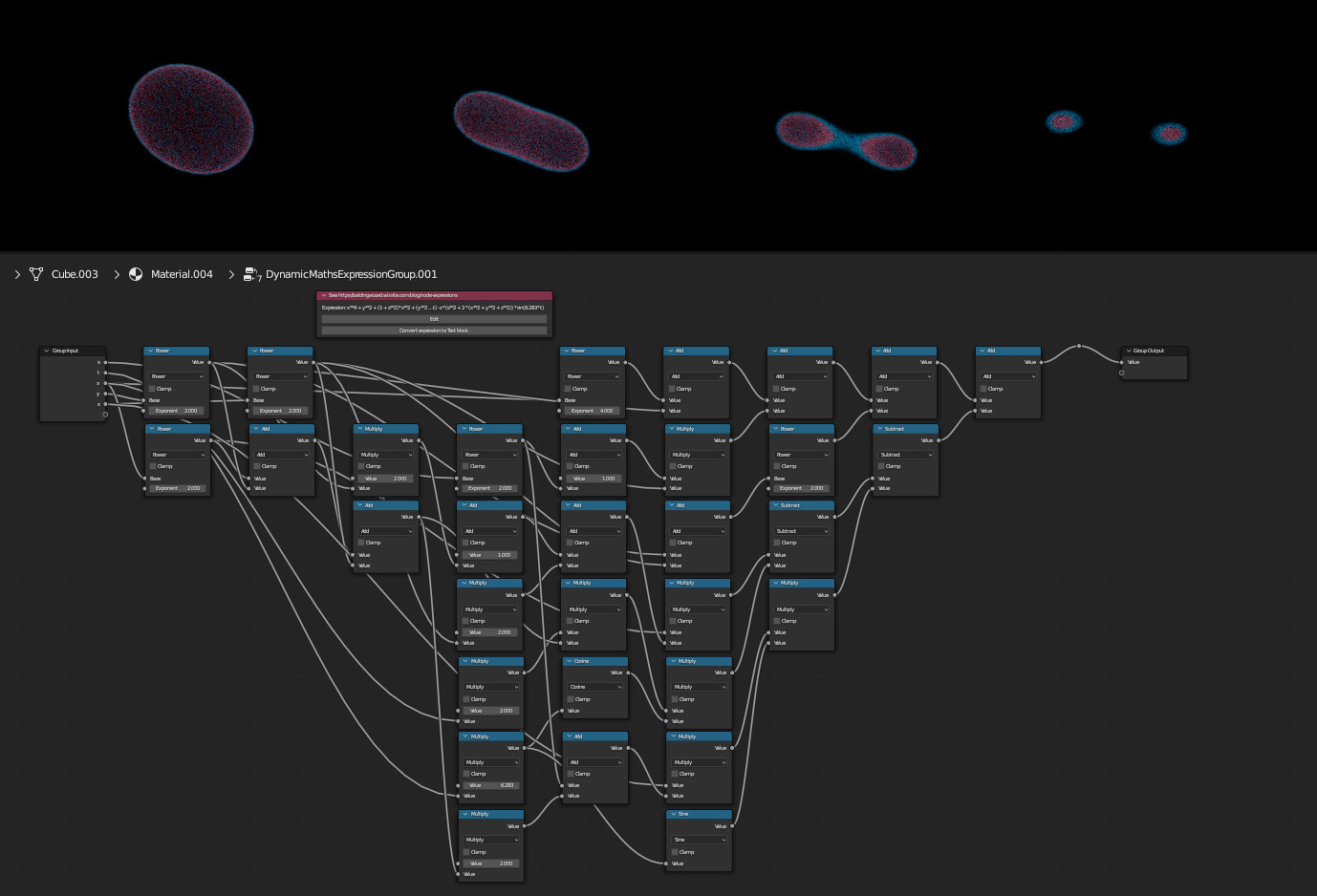}
    \caption{Determinant values in real time volume shading, and the corresponding shader nodes of a fuzzy two-nodes graph.}
    \label{fig:nodetree}
\end{figure}

Another way to visualize zero mode surfaces based on graphs is to just use Blender and its render engine for plotting. Together with an addon \cite{Sedman:2018} to render the determinant as volumetric density. The density is high for determinant values near zero and rapidly decreases within a short range. This highlights the regions where the determinant is positive or negative and shows how these values change. We examine the rotating edge of two nodes. In Mathematica, the zero-mode surface appears as follows
\begin{figure}[H]
    \centering
    \includegraphics[width=1.0\linewidth]{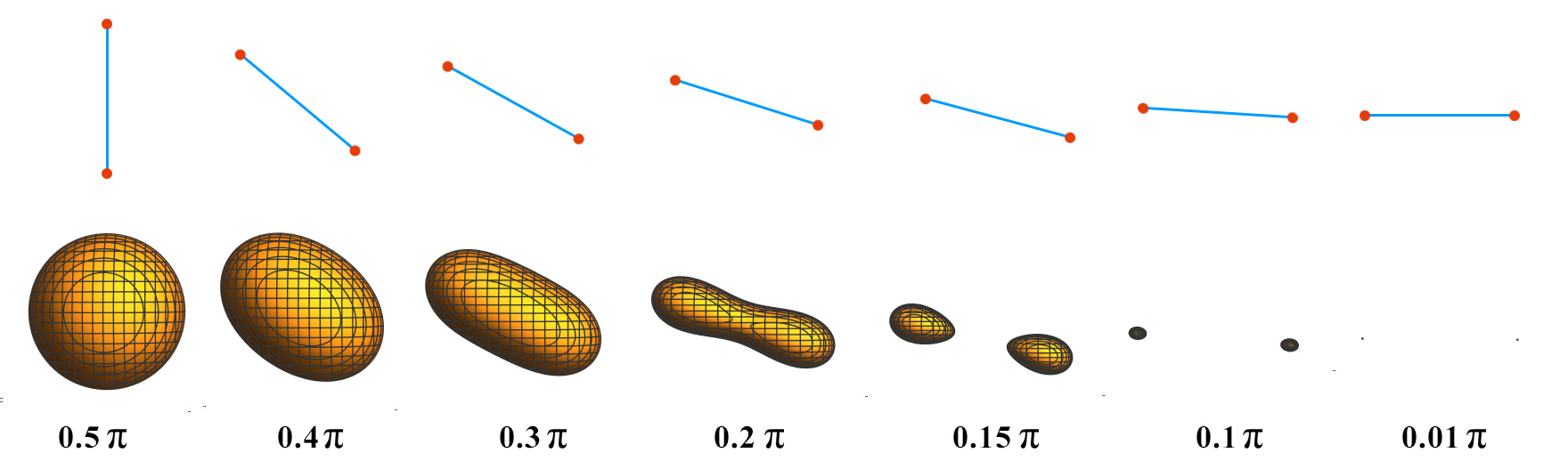}
    \caption{Turning a simple graph }
    \label{fig:rotatestick}
\end{figure}
We see how the surface shrinks to two points. A parameter $\alpha$ drives the rotation of the edge with the following matrices
\begin{eqnarray} 
    X=\left(\begin{array}{cc}\cos (\alpha) & 1 \\   1 & 0 \end{array}\right),&  Y=\left(\begin{array}{cc} 0 & -i \\ i & 0\end{array}\right),&
    Z=\left(\begin{array}{cc}\sin (\alpha) & 0 \\ 0 & 0\end{array}\right).
\end{eqnarray}
These matrices were turned to a volumetric shader and rendered in Blender's 3D viewport. In Figure \ref{fig:nodetree}, from left to right, we see that the greater the volume of the surface of the zero mode, the larger the $Z$ difference between the nodes. And with that the fuzzy surfaces grow.
The red color indicates negative values of the determinant, and the blue color indicates positive values. The determinant separates into two negative areas that vanish, as the nodes' $z$ coordinates coincide.
\\
Notice that the thicknesses of these layers change. The red and blue layers mark the interval $[-0.1, 0.1]$ for the determinant values at each point in space. Increased thickness means the determinants values spatially change less. The magnitude of the determinants gradient decreases with horizontal edges.
Blender's volume shading offers a broad range of tools to investigate specific properties of our determinants. This workflow seizes parallelized GPU architecture and lets zero mode surfaces of simple graphs be visualized in realtime.

\clearpage

\section{Parameter-dependent examples}
For the next examples, we did not construct the graphs using Blender, instead, we manually wrote the matrices in Mathematica, in order to implement parameters that can be animated for dynamic contourplots. 
\subsection{From cylinder to torus}
\begin{figure}[H]
    \centering
    \includegraphics[width=0.7\linewidth]{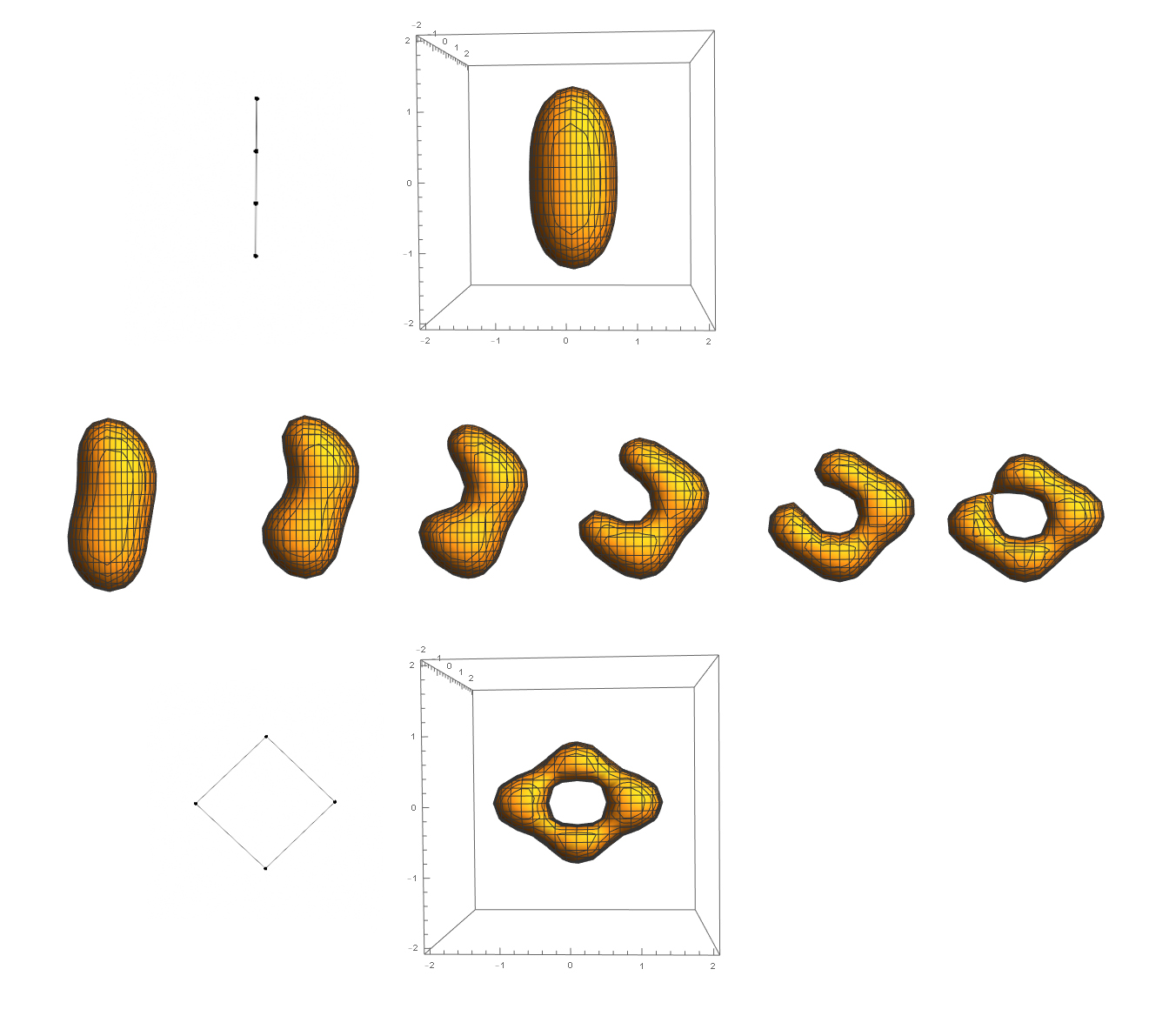}
    \caption{From cylinder to torus.}
    \label{fig:sphere-torus}
\end{figure}

We explore the transition from a fuzzy cylinder to a fuzzy torus. First, we define two matrices: one for the fuzzy cylinder $H_C$ and one for the fuzzy torus $H_T$, both with the same number of nodes. A plot can be animated using a parameter $p\in[0, 1]$ . This parameter combines the two matrices into one, $ H(p)=(1-p)\cdot H_C+p\cdot H_T$. 
The result is a transition with fuzzy surfaces for the states in-between.
 The plots in Figure $\ref{fig:sphere-torus}$ were made, using the following matrices $X^1_T$, $X^2_T$, $X^3_T$ for the torus and $X^1_C$, $X^2_C$, $X^3_C$ for the cylinder, with $s_y=-is_x=-\frac{i}{2}$
 \begin{center}
\begin{minipage}{\linewidth}
\small
\[
X^1_T = \begin{pmatrix}
0 & s_x & s_x & 0 \\
s_x & -1 & 0 & s_x \\
s_x & 0 & 1 & s_x \\
0 & s_x & s_x & 0
\end{pmatrix}
,
X^2_T = \begin{pmatrix}
0 & s_y & s_y & 0 \\
\bar{s}_y & 0 & 0 & s_y \\
\bar{s}_y & 0 & 0 & s_y \\
0 & \bar{s}_y & \bar{s}_y & 0
\end{pmatrix}
,
X^3_T = \begin{pmatrix}
1 & 0 & 0 & 0 \\
0 & 0 & 0 & 0 \\
0 & 0 & 0 & 0 \\
0 & 0 & 0 & -1
\end{pmatrix}
\]
\begin{equation}
X^1_C = \begin{pmatrix}
0 & s_x & 0 & 0 \\
s_x & 0 & s_x & 0 \\
0 & s_x & 0 & s_x \\
0 & 0 & s_x & 0
\end{pmatrix}
,
X^2_C = \begin{pmatrix}
0 & s_y & 0 & 0 \\
\bar{s}_y & 0 & s_y & 0 \\
0 & \bar{s}_y & 0 & s_y \\
0 & 0 & \bar{s}_y & 0
\end{pmatrix}
,
X^3_C = \begin{pmatrix}
\frac{3}{2} & 0 & 0 & 0 \\
0 & \frac{1}{2} & 0 & 0 \\
0 & 0 & \frac{-1}{2} & 0 \\
0 & 0 & 0 & \frac{-3}{2}
\end{pmatrix}
\\
\end{equation}
\end{minipage}
 \end{center}
%\clearpage
So that
\[
H_T=\begin{pmatrix}
    X^3_T-z & X^1_T-x+ i(X^2_T-y)\\
    X^1_T-x+i(X^2_T-y) & -(X^3_T-z)
 \end{pmatrix}
\]
\\
and
\begin{equation}
H_C=\begin{pmatrix}
    X^3_C-z & X^1_C-x+ i(X^2_C-y)\\
    X^1_C-x+i(X^2_C-y) & -(X^3_C-z)
 \end{pmatrix}.
\end{equation}
\\
 Figure $\ref{fig:sphere-torus}$ shows the contour plot for $det(H(p))=0$.
%This demonstrates that the linear combination of two graphs that have a zero-surface? also generates zero mode surfaces.\\%
Note that it leads to the same result if one first combines the matrices $X^a_C$ and $X^a_T$ to a convex linear combination $X^a(p)=(p-1)X^a_C + pX^a_T$ and then computes the matrix $H(p)$, as if one first forms the $H_C,H_T$ matrices of the individual fuzzy spaces and then combines them into a convex linear combination $H(p) = (1-p)\cdot H_C+p\cdot H_T$.
One remarkable observation is the transition of topology from genus 0 to genus 1. As shown in Figure $\ref{fig:sphere-torus}$, the cylinder first closes into a torus at a single point where the deformed ends touch. As shown in in Figure $\ref{fig:sphere-torus}$, the first time, that the cylinder closes into a torus, is achieved by one single point where the pointed ends of the deformed cylinder touch.

\subsection{Separation of a zero-manifold}
\begin{figure}[H]
    \centering
    \includegraphics[width=1.0\linewidth]{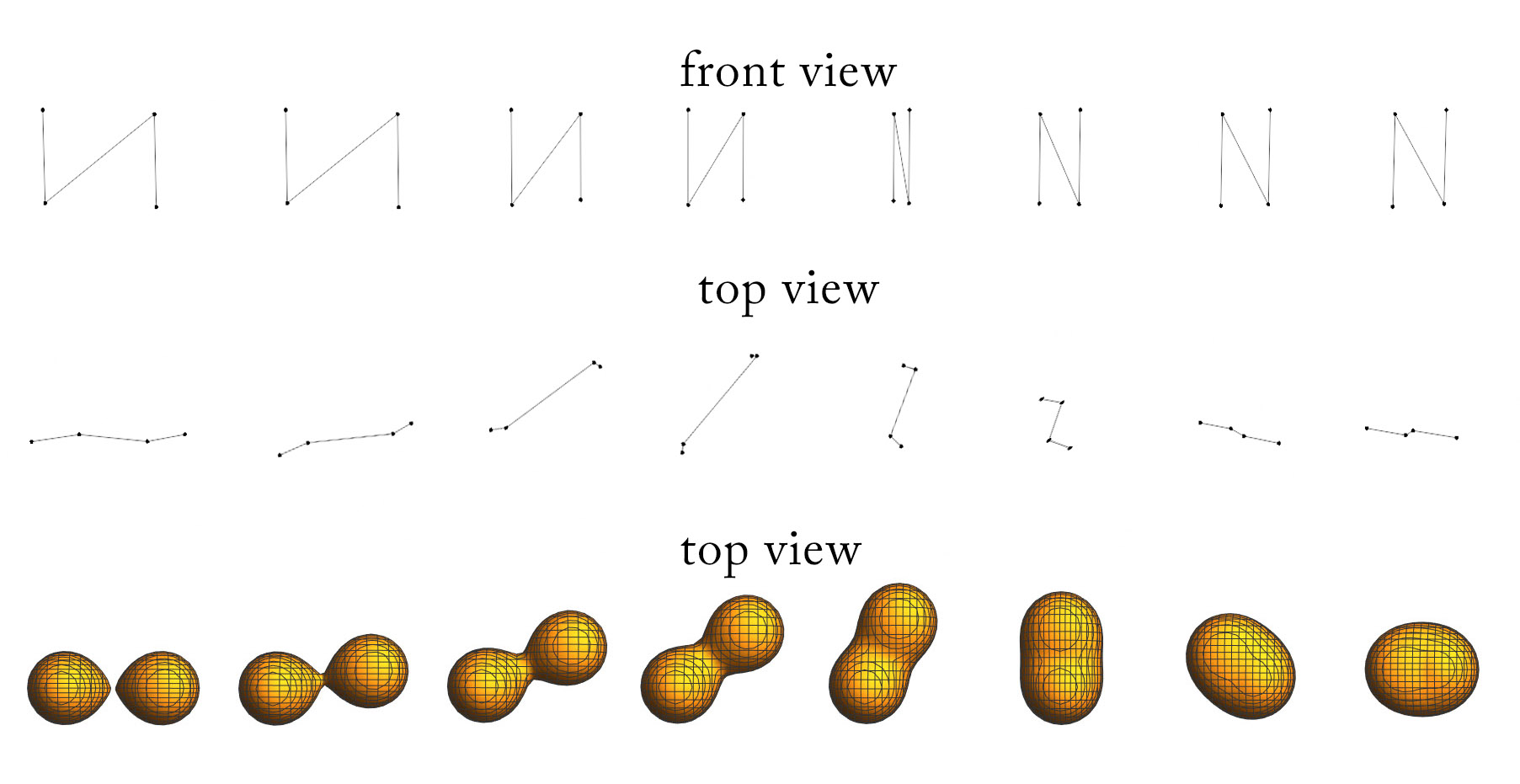}
    \caption{Two orbiting vertical edges, connected with a diagonal edge.}
    \label{fig:time_dep}
\end{figure}

 We build a graph with a variable parameter, to examine the transition between separated and connected manifolds. For this we use a fuzzy sphere orbiting another fuzzy sphere along an elliptical trajectory. The two fuzzy spheres are connected with a third edge. This third edge goes from one's lower to the other's upper node, to avoid horizontal edges:
\[
\text{{fcos}} = \frac{1}{2} + 2 \cos(a); \quad \text{{fsin}} = 2 \sin(a); \quad s = 1;
\]
\small
\begin{minipage}{\textwidth}
\begin{equation}
\text{X} = 
\begin{pmatrix}
    \frac{-1}{2}  & s & 0 & s \\
    s &  \frac{-1}{2}  & 0 & 0 \\
    0 & 0 & \text{fcos}  & s \\
    s & 0 & s & \text{fcos} 
\end{pmatrix}
,
\text{Y} = 
\begin{pmatrix}
    0  & -i s & 0 & -i s \\
    i s & 0  & 0 & 0 \\
    0 & 0 & \text{fsin}  & -i s \\
    i s & 0 & i s & \text{fsin} 
\end{pmatrix}
,
\text{Z} = 
\begin{pmatrix}
    1  & 0 & 0 & 0 \\
    0 & -1  & 0 & 0 \\
    0 & 0 & 1  & 0 \\
    0 & 0 & 0 & -
    1 
\end{pmatrix}.
\end{equation}
\\
\end{minipage}
\\
As shown in Figure  $\ref{fig:time_dep}$, when the two spheres approach each other, a 'droplet-like' connection forms between them.
\clearpage
\subsection{Visualizing deformations and gauge theory}
As it is standard procedure in non-commutative field theory, we deform the coordinates 
$X^{a}$ with a gauge field $A^{a}$. We investigate the effect of this deformation on both graphs and zero mode surfaces through the analysis of two simple examples.\\

\begin{figure}[H]
    \centering
    \includegraphics[width=0.5\linewidth]{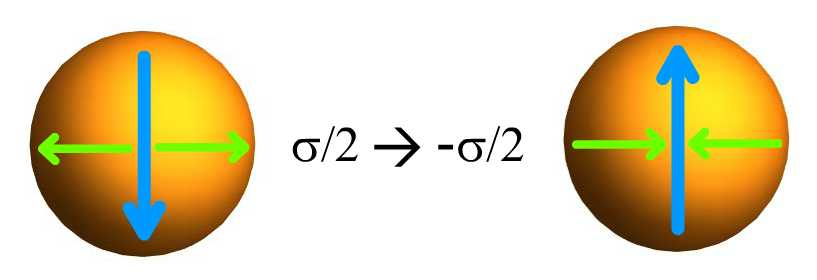}
    \caption{Gauge deformations change the graph data.}
    \label{fig:gaugegraphdata}
\end{figure}

\paragraph{Zero curvature deformations.}
Let us consider the deformation of the coordinates $X^{a}$ by a gauge field $A^{a}$
\begin{eqnarray}
\tilde{X}^{a}=X^{a}+A^{a}.
\end{eqnarray}
 The commutator of these deformed coordinates reads
\begin{eqnarray}
[\tilde{X}^{a},\tilde{X}^{b}]=[X^{a},X^{b}]+[X^{a},A^{b}]-[X^{b},A^{a}]+[A^{a},A^{b}]
\end{eqnarray}
this can be written as deformed commutator
\begin{eqnarray}
\tilde{\theta}^{ab}=\theta^{ab}+F^{ab},
\end{eqnarray}
with the field strength or curvature 
\begin{eqnarray}\label{curv}
F^{ab}=[X^{a},A^{b}]-[X^{b},A^{a}]+[A^{a},A^{b}].
\end{eqnarray}
For a clear example, we consider the simplest fuzzy sphere using the $2\times2$ matrices
\begin{eqnarray}
X = \frac{1}{2}\left( \begin{array}{cc}
   0 & 1 \\
     1&  0\\
       \end{array} \right)=\frac{1}{2}\sigma^{1},
       Y = \frac{i}{2}\left( \begin{array}{cc}
   0& -1 \\
    1&  0\\
       \end{array} \right)=\frac{1}{2}\sigma^{2},
       Z = \frac{1}{2}\left( \begin{array}{cc}
  1 & 0 \\
     0& -1\\
       \end{array} \right)=\frac{1}{2}\sigma^{3}\nonumber\\
\end{eqnarray}
and deform this matrices by a gauge field $A^{a}=A^{a}_{\alpha}\sigma^{\alpha}$, where $\sigma^{\al}$ are the Pauli matrices with the commutator relations $[\sigma^\alpha,\sigma^\beta]=2i\varepsilon^{\alpha\beta\gamma}\sigma^\gamma$. If we plug this in the equation  ($\ref{curv}$) we find
\begin{equation}
F^{ab}=i(A^{b}_{\al}\varepsilon^{a\al\ga}-A^{a}_{\al}\varepsilon^{b\al\ga}+A^{a}_{\al}A^{b}_{\be}\varepsilon^{\al\be \ga})\sigma^{\ga}=F^{ab}_{\ga}\sigma^{\ga}
\end{equation} and in components
\begin{equation}
F^{ab}_1=\begin{pmatrix}
0 & -A^1_3+2A^1_2 A^2_3-2A^1_3A^2_2 & A^1_2+2A^1_2A^3_3-2A^1_3A^3_2 \\
. & 0 & A^3_3+A^2_2+2A^2_2 A^2_3-2A^2_3A^2_2 \\
. & . & 0 \\
\end{pmatrix}\nonumber
\end{equation}
\begin{equation}
F^{ab}_2=\begin{pmatrix}
0 & A^2_3-2A^1_1 A^2_3+2A^1_3A^2_1 & -A^3_3-A^1_1 -2A^1_1A^3_3+2A^1_3A^3_1 \\
. & 0 & -A^2_1-2A^2_1A^3_3+2A^2_3A^3_1 \\
. & . & 0 \\
\end{pmatrix}\nonumber
\end{equation}
\begin{equation}
F^{ab}_3=\begin{pmatrix}
0 & A^2_2+A^1_1+ 2A^1_1A^2_2-2A^1_2A^2_1 & A^3_2+2A^1_1A^3_2-2A^1_2A^3_1 \\
. & 0 &-A^3_1+2A^2_1A^3_2-2A^2_2A^3_1 \\
. & . & 0 \\
\end{pmatrix},
\end{equation}\\
where the dots in  the matrices indicate the negative transposed upper triangular matrix entries.
For zero curvature $F^{ab}_{\ga}=0$ this can be solved by setting 
 $A^{a}_{\al}=-1$ when  $a=\al$ and $A^{a}_{\al}=0$ for all other cases.
This leads to the gauge field
\begin{equation}
A^{a}=-\sigma^{a}
\end{equation}
and the deformed coordinates are
\begin{eqnarray}
\tilde{X}^{a}=X^{a}+A^{a}=\frac{\sigma^a}{2}-\sigma^a=-\frac{\sigma^a}{2}=-X^a.
\end{eqnarray}
\\ 
Note, that the coordinates $\tilde{X}^{a}$ still satisfy the relation $\sum( \tilde{X}^{a})^2=3/4$ and define the same zero mode surface as the coordinates $X^a$ do.
But on the level of graphs the radii become negative and the $z$-coordinates are exchanged, as shown in Figure $\ref{fig:gaugegraphdata}$.
%The coordinates $\tilde{X}^{a}$ define the same zero mode surface as the coordinates $X^a$ and the same fuzzy sphere. But on the graph the radii become negative, and the $z$-coordinates are exchanged, as shown in Figure $\ref{fig:gaugegraphdata}$

The actual difference between the two fuzzy spheres lies in their Lie algebras.
The matrices $\tilde{X}^{a}$ satisfy the commutation relations $[\tilde{X}^{a},\tilde{X}^{b}]=if^{abc}\tilde{X}^{c}$ with the structure constants $f^{abc}=-\varepsilon^{abc}$, unlike the matrices  $X^a$ satisfying the commutation relations  $[X^{a},X^{b}]=i\varepsilon^{abc}X^{c}$.
%But this actually leads to different structure constants in the Lie algebra $%[\tilde{X}^{a},\tilde{X}^{b}]=if^{abc}\tilde{X}^{c}$ with $f^{abc}=-\varepsilon^{abc}$.

\begin{figure}[H]
    \centering
    \includegraphics[width=1.0\linewidth]{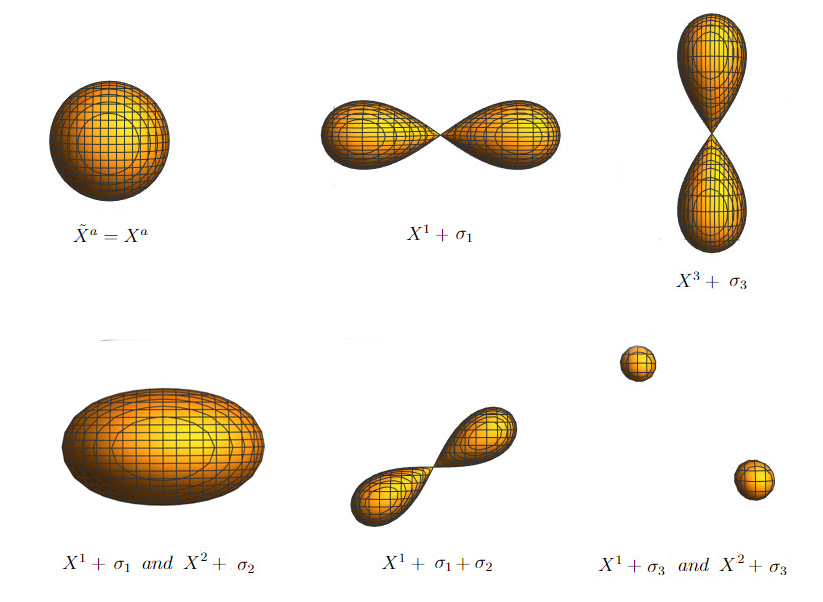}
    \caption{Deformations of the fuzzy sphere.}
    \label{fig:gauge}
\end{figure}

\paragraph{More general deformations.}
We stay with the example of the smallest fuzzy sphere as above. We deform the coordinates $X^a$ again with the gauge field $A^{a}=A^{a}_{\alpha}\sigma^{\alpha}$, but now we do not restrict the deformation to zero curvature. We only show in the plots Figure \ref{fig:gauge} some specific values of the coefficients $A^{a}_{\alpha}$ of the gauge field. 

In Figure \ref{fig:gauge} the top left is the undeformed fuzzy sphere. The top-mid-plot shows a deformation of the $s_{x}$ radius in the $X^{1}$ matrix. Note that this deformation still leads to a Lie algebra with modified structure constants given by the undeformed fuzzy sphere. Geometrically, this deformation describes a spheroid, but the zero-mode surface appears in the plot as a singular surface. On the top right plot, the coordinates of the graphs nodes are deformed, rather than the radii. The zero-mode surface appears singular as in the previous case, similarly to the top middle plot, but in this case the deformation acts on the z coordinates.
 \\In the plot on the bottom left, it is noticeable that when the radii are deformed to the same order of magnitude, the zero-mode surface shapes a spheroid. Geometrically, the deformed matrices also shape a spheroid. In the plots at the bottom middle and bottom right, we show arbitrary deformations, which in general do not lead to a fuzzy ellipsoid. However, the resulting zero surfaces are interesting in their own right. To see the dynamic influence of the deformations on the zero-mode surface, we recommend running the Mathematica notebook ''Fuzzy Sphere Deformations.nb''  \cite{SchSchM1:2024}.

\section{Conclusions and outlook}
This paper investigates methods of matrix models and fuzzy geometry to represent and analyze fuzzy surfaces and their zero-mode surfaces, with a focus on complex geometrical constructs such as the two-dimensional fuzzy Trefoil knot figure \ref{fig:TREFOIL}. Using the 3D modeling software Blender and the computational tool Wolfram Mathematica, we established a comprehensive workflow to create and analyze intricate graphs. The methodology involves encoding coordinates and connectivity of graphs into matrices, allowing for precise visualization and manipulation of fuzzy surfaces and zero-mode surfaces.\\ \\
Our findings emphasize the critical role of edge direction in maintaining the structural integrity of zero-mode surfaces. Altering the direction of an edge can greatly affect the shape of the surface, as seen in the transition from a spherical to an hourglass configuration. Therefore, we avoided horizontal edges to reduce gaps between nodes, which can lead to discontinuities in the surface.\\
It is notable that for highly symmetric fuzzy spaces, the commutator of the matrices plays a very central role in non-commutative geometry. In our fuzzy space examples, the commutator is not computed. In some cases, we have not explicitly written the matrices corresponding to the graphs derived from Blender. We provide all the Mathematica notebooks and the Blender files for download via links listed in section 6. The procedure is more numerical in nature, and the commutator or the explicit form of the matrices is not crucial for the graphical representation of the zero-mode surfaces. Furthermore, the polynomials in the variables $x,y,z$ given by the zero-mode surfaces deserve further analysis.\\ \\
A major result of the paper is the process of modeling the fuzzy Trefoil knot, balancing the complexity of the graph with the need for a detailed representation. The resulting 30-node Trefoil knot graph exemplifies the effectiveness of our approach in capturing the essential features of complex geometrical shapes. Naturally, more complex fuzzy knots can, in principle, be constructed using our method. An alternative workflow is presented, using Blender's volume shading to leverage parallel GPU architecture for faster visualization. \\
We also examined the interpolation between a fuzzy cylinder and a fuzzy torus, observing how the topology transitions from one form to another in the zero-mode surface. There is a point where the cylinder closes to become a torus, and this process should be studied further. \\
We investigated the behavior of the zero-mode surface by continuously altering graphs and matrices through the use of parameters. For example, as two fuzzy spheres approach each other, a "droplet-like" structure forms, varying with the distance between the connected fuzzy spheres. This behavior is similar to "Metaballs" in 3D software. An alternative workflow is presented, using Blender's volume shading to leverage parallel GPU architecture for faster visualization. In our latest results, we consider simple examples where the coordinates of the simplest fuzzy sphere are deformed, corresponding to switching on a gauge field in non-commutative geometry. These examples demonstrate the effect of deformation on the level of the fuzzy surface, showing how the graph changes through deformation and how the resulting zero surface changes accordingly.\\
The tools and techniques developed in this study provide a robust framework for visualizing and analyzing fuzzy geometries. Our workflow, which combines intuitive graph modeling with advanced mathematical plotting, facilitates the exploration of fuzzy spaces and their properties. This research lays the groundwork for future studies aimed at refining these methods and applying them to more advanced geometrical models, and even to 3D algorithms and physical models, which we will investigate in the future.
It would be interesting to investigate the fuzzy surfaces described above using the methods developed in \cite{Shimada:2003ks}. This involves examining the sequence of eigenvalues of the matrices, which has a connection to Morse theory. Particularly for the Trefoil knot, there are possible relations to knot data or methods to extract knot data from the matrices $X,Y,Z$ or from the zero mode surface polynomial. It also raises the question of how to extend complex fuzzy surfaces, such as the fuzzy Trefoil knot, using methods from \cite{Sykora:2017} to arbitrary $N\times N$ matrices with a classical limit.\\ \\
In summary, this paper contributes to the field by providing practical insights and methodologies for representing and analyzing fuzzy surfaces. The findings and techniques described here could greatly advance the study of fuzzy spaces and their applications in quantum physics and fuzzy geometry.
\\
\clearpage
\section{Appendix}
\paragraph{Blender to Mathematica Python script.}
The Python package sympy must be installed before running the code.
Here is a flowchart of the script that reads the modeled graphs:
\begin{figure}[H]
    \centering
    \includegraphics[width=1\linewidth]{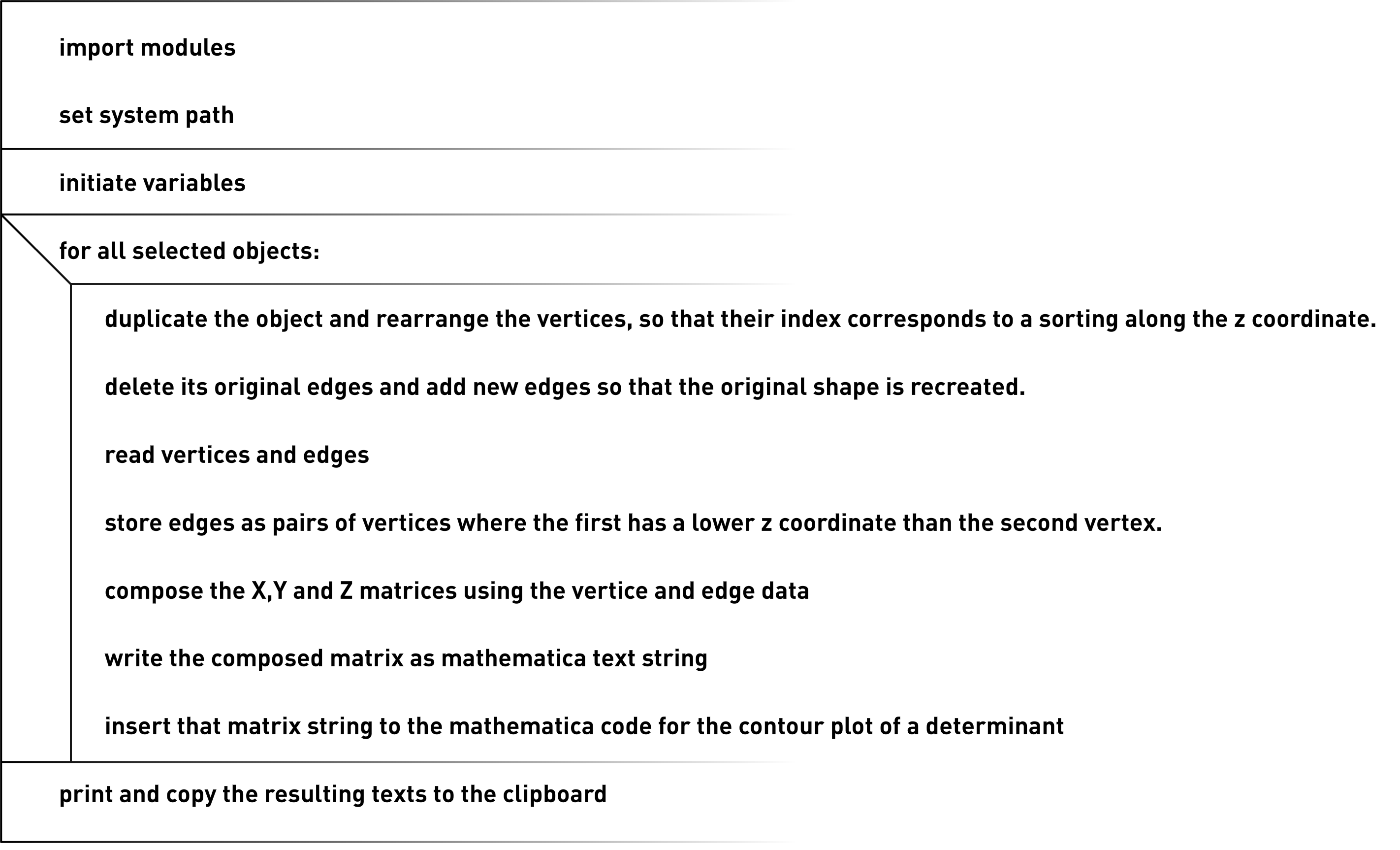}
    \caption{Python script flow chart.}
    \label{fig:enter-label}
\end{figure}
You can download the Python file with the code here \cite{SchSchB1:2024} and notebooks in Mathematica here
\cite{SchSchM1:2024}.

\end{document}